\documentclass{article}
\usepackage{ijcai23}

\usepackage{times}
\usepackage{soul}
\usepackage{url}
\usepackage[hidelinks]{hyperref}
\usepackage[utf8]{inputenc}
\usepackage[small]{caption}
\usepackage{graphicx}
\usepackage{amsmath}
\usepackage{amsthm}
\usepackage{booktabs}
\usepackage{algorithm}
\usepackage{algorithmic}
\usepackage[switch]{lineno}
\usepackage{amsfonts}

\title{Towards Robust GAN-generated Image Detection: a Multi-view Completion Representation}

\author{
Chi Liu$^1$
\and
Tianqing Zhu$^1$\and
Sheng Shen$^{2}$\And
Wanlei Zhou$^3$\\
\affiliations
$^1$School of Computer Science, University of Technology Sydney, Australia\\
$^2$School of Electrical and Information Engineering, The University of Sydney, Australia\\
$^3$City University of Macau, Macao SAR, China \\
\emails
tianqing.zhu@uts.edu.au
}

\begin{document}

\maketitle

\begin{abstract}
   GAN-generated image detection now becomes the first line of defense against the malicious uses of machine-synthesized image manipulations such as deepfakes. Although some existing detectors work well in detecting clean, known GAN samples, their success is largely attributable to overfitting unstable features such as frequency artifacts, which will cause failures when facing unknown GANs or perturbation attacks. To overcome the issue, we propose a robust detection framework based on a novel multi-view image completion representation. The framework first learns various view-to-image tasks to model the diverse distributions of genuine images. Frequency-irrelevant features can be represented from the distributional discrepancies characterized by the completion models, which are stable, generalized, and robust for detecting unknown fake patterns. Then, a multi-view classification is devised with elaborated intra- and inter-view learning strategies to enhance view-specific feature representation and cross-view feature aggregation, respectively. We evaluated the generalization ability of our framework across six popular GANs at different resolutions and its robustness against a broad range of perturbation attacks. The results confirm our method's improved effectiveness, generalization, and robustness over various baselines. 
\end{abstract}

\section{Introduction}

\noindent AI-powered image manipulation techniques, such as deepfakes, are constantly evolving thanks to the continuous advances in deep generative models, particularly generative adversarial networks (GANs) \cite{goodfellow2014generative}. The quality and fidelity of the generated images have reached a photorealistic level that is indistinguishable from real images to human eyes. Alongside the technical advance, society is raising significant concerns regarding the abuse of these techniques to create and spread misleading information, which will cause a trust crisis where ``seeing is no longer believing." To tackle the issues, the research community has been dedicated to developing powerful forensics tools against malicious image manipulations. One crucial and promising direction is detecting GAN-generated fake images, considering the ubiquitous adoptions of GANs in image manipulation tasks. 

\begin{figure}
    \centering
    \includegraphics[width=0.48\textwidth]{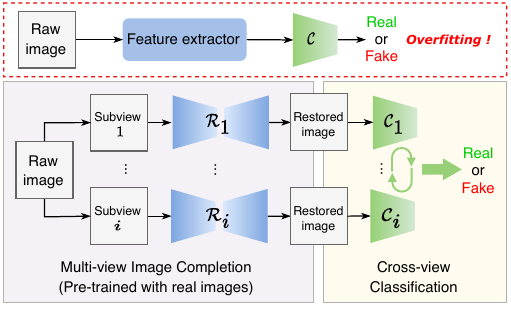}
    \caption{Instead of learning GAN-specific features directly from fake images which may lead to overfitting, our framework incorporates multi-view completion and classification to model diverse distributional discrepancies between real and fake images, which can generalize to unknown fake patterns. $\mathcal{R}$: Restorer; $\mathcal{C}$: Classifier.}
    \label{fig:intro}
\end{figure}

Most detection methods typically train CNN classifiers to learn specific features to distinguish GAN-generated images from real ones \cite{hu2021exposing,liu2020global,marra2019gans,dzanic2020fourier,frank2020leveraging,durall2020watch}, which work satisfactorily on clean test samples from the same GANs used in training. However, their performances will decrease dramatically when facing samples generated by unknown GANs or perturbed by noises, leading to limited practical reliability. One primary reason is that a deep CNN classifier may easily overfit unstable GAN-specific features of the training samples, particularly the low-level frequency-domain artifacts \cite{wang2020cnn,he2021beyond,jeong2022bihpf,jeong2022frepgan}. Previous studies have proved that conspicuous artifacts exist in the spectra of GAN-generated images \cite{wang2020cnn,frank2020leveraging}: Despite being easily identified by classifiers, these artifact patterns are inconsistent, varying significantly among different GAN models or perturbations. As a result, the classifier overfitting a specific frequency pattern will suffer from weak generalization ability and robustness in detecting other frequency patterns.

Based on the understanding of the overfitting issue, we are motivated to design an improved detection model with two requirements: (1) reduce the dependency on unstable low-level frequency features; and (2) learn a robust feature representation from other types of information, such as regional consistency, and color or textural details of images. Instead of directly learning detectable features from fake images, which potentially leads to the frequency overfitting problem, we propose a novel detection framework that incorporates a multi-view image completion learning and a cross-view classification learning processes, as sketched in Fig. \ref{fig:intro}. The framework can learn a strong and stable feature representation from diverse frequency-independent, view-specific information, resulting in outperforming generalization and robustness when facing unknown GANs or perturbations. 

In the multi-view completion process, multiple view-to-image completion models are learned \textit{with real images only}, and then used to characterize diverse distributional discrepancies between real and fake images. In contrast to overfitting specific GAN patterns, the compact distributions of the image-missing characteristics modeled from real images are more likely to distinguish unknown, out-of-distribution fake images from real ones \cite{ruff2020deep}. In addition, the view-to-image completions can align the frequency patterns of different types of fake samples with that of real images, which helps reduce the classifier's frequency bias. Then, in the cross-view classification, the real and fake samples synthesized from each incomplete view are fed into an independent classifier. The multi-scale feature concatenation and low-pass residual-guided attention modules are devised to strengthen the intra-view feature representation. The independent classifiers are finally combined using an adaptive loss fusion strategy to enhance the learning from cross-view information. Our contributions are highlighted as follows:

\begin{itemize}
    \item We propose a novel GAN-generated image detection framework using multi-view completion classification learning to build a robust feature representation for detecting unknown GANs and perturbations. 
    \item We devise several novel modules and learning strategies that effectively benefit the framework's ability to capture and incorporate diverse view-specific features.  
    \item We perform extensive evaluations which validate the significantly improved generalization and robustness of our framework in a wide range of settings varying in image resolutions, GAN types, and perturbation methods. 
\end{itemize}

\section{Generated Image Detection: A Review}
\paragraph{Image-domain detection.} Image-domain detection extracts detectable traces from the pixel inputs. Earlier works tended to train a CNN to learn deep features in a data-driven manner \cite{marra2018detection,tariq2018detecting}, while more recent works prefer to craft specific features for higher detection accuracy, such as the co-occurrence matrices \cite{nataraj2019detecting,barni2020cnn}, saturation \cite{mccloskey2019detecting}, specular highlights \cite{hu2021exposing} and texture cues \cite{liu2020global}. \cite{marra2019gans} and \cite{yu2019attributing} pointed out that a GAN will leave a unique fingerprint containing the source model information in the generated images. Some other works improve on the network design, where novel learning strategies or modules, such as incremental learning \cite{marra2019incremental}, self-attention mechanism \cite{jeon2020fdftnet,mi2020gan} and vision transformer \cite{wang2022m2tr}, are adopted. 

\paragraph{Frequency-domain detection.} 
Frequency-domain detection relies on identifying the frequency discrepancy between GAN-generated images and real images \cite{dzanic2020fourier,frank2020leveraging,durall2020watch}. Frequency discrepancy can be easily captured in various spectral representations by a classifier. For example, \cite{frank2020leveraging} found that even a shallow CNN can achieve a high detection accuracy using the 2D Discrete Cosine Transform (DCT) coefficients as input data. \cite{qian2020thinking} proposed a dual-branch network that extracts the global and local DCT features. \cite{dzanic2020fourier} and \cite{durall2020watch} proposed to transform the 2D Fast Fourier Transform (FFT) magnitude into 1D power profile as detectable features. \cite{liu2021spatial} found that more distinguishable features can be extracted in the phase spectrum than in the amplitude spectrum. Unfortunately, some recent studies also pointed out that the frequency features are unstable and easy to be concealed \cite{dzanic2020fourier,durall2020watch,liu2022making,huang2020fakepolisher,jung2021spectral,dong2022think}. Thus, detectors heavily relying on frequency features are vulnerable and weakly generalized. 

\paragraph{Generalized and robust detection.}
Generalized and robust detection of GAN-generated images is now in high demand. Most existing works involve a preprocessing operation to strengthen the representation of generalized and robust features. \cite{zhang2019detecting} pointed out that a detector can generalize between two GANs with similar spectral artifacts in their generated images, which in turn confirms that a detector is likely to overfit specific frequency patterns. \cite{wang2020cnn} explored the effects of different data augmentation strategies such as compression and blurring in improving a detector's generalization ability. \cite{jeong2022bihpf} proposed to preprocess the fake images with a bilateral high-Pass filter, which amplifies the effect of the common frequency-level artifacts shared by different GANs. \cite{jeong2022frepgan} designed a frequency-level perturbation framework to erode the GAN-specific spectral artifacts in generated images before feeding them to a detector. \cite{he2021beyond} proposed to re-synthesize training images using a super-resolution model pre-trained with real images to help extract robust features and isolate fake images. 

\begin{figure*}[htb]
    \centering
    \includegraphics[width=\textwidth]{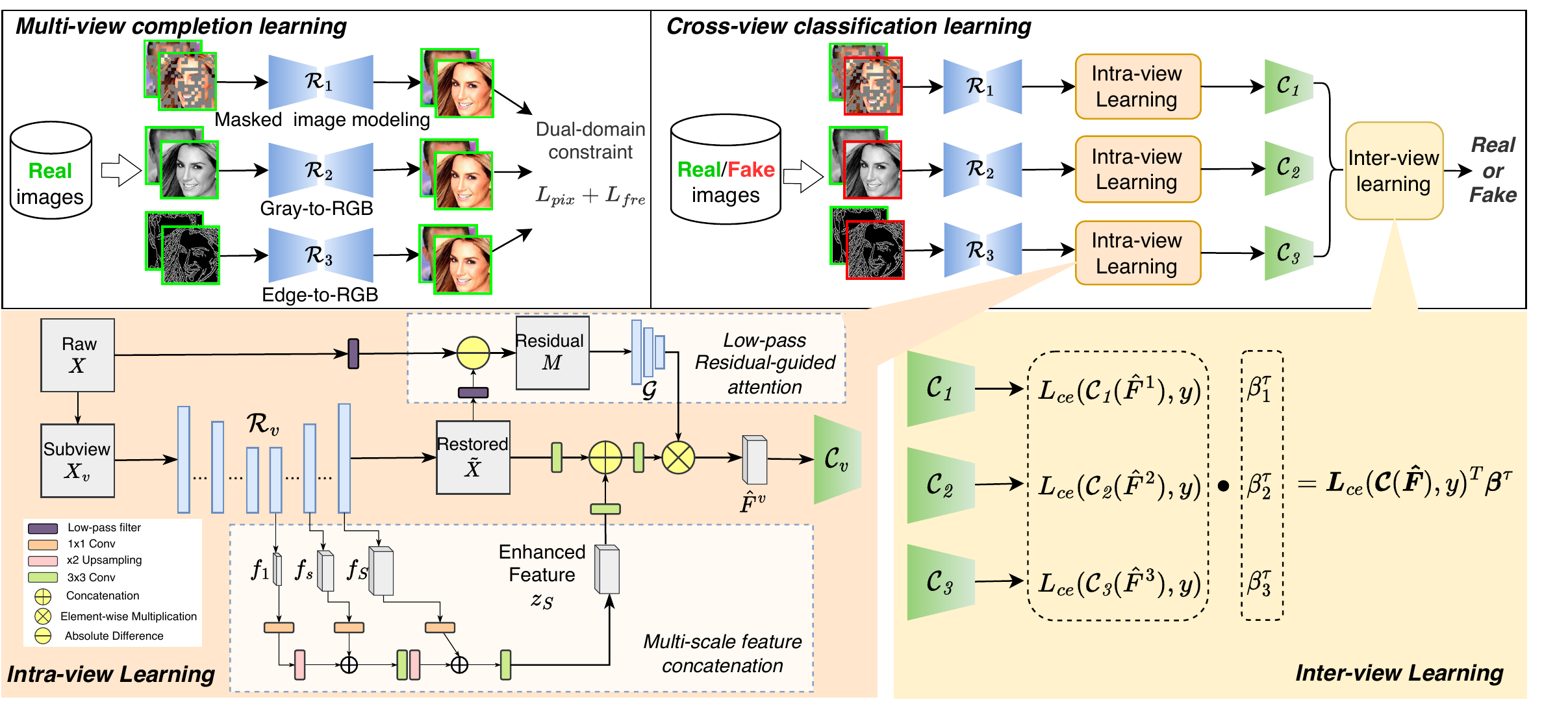}
    \caption{The overview of our framework (white box). Several restorers first learn different distributions of real images via multi-view completion learning. Then for each view, a classifier captures the view-specific distributional discrepancy between real and fake images via intra-view learning. The low-pass residual-guided attention and multi-scale feature concatenation modules are devised to strengthen intra-view learning (orange box). All base classifiers are finally fused to perform inter-view learning for robust detection (yellow box).}
    \label{fig:overview}
\end{figure*}

\section{The Proposed Framework}
We design the \textbf{M}ulti-view \textbf{C}ompletion \textbf{C}lassification \textbf{L}earning (MCCL) to build a novel multi-view, frequency-independent feature representation for robust detection of GAN-generated images. As shown in Figure \ref{fig:overview}, the framework jointly trains a set of restorers and classifiers. The restorers are trained with real images only, and each learns to reconstruct the full image from one particular incomplete view. Then, both real and fake images are processed by each restorer through the same view-to-image pipeline. Since the recovery of missing information is governed by real images' characteristics only, the distributional difference between the reconstructed real and fake samples can be reflected in the restored information. Then, for each view, a classifier is trained based on the reconstructed samples to capture the view-specific distributional discrepancy. We combine the multi-scale features encoded by different decoding layers of each restorer with the restored image as the classifier's input. A low-pass residual-guided attention module is employed at the entry of the classifier to highlight the reconstruction difference between real and fake images. A self-adaptive loss fusion module is additionally designed to combine the decisions of multiple classifiers to facilitate inter-view learning.

\subsection{Multi-View Image Completion Learning}
Several independent encoder-decoder-based restorers $\boldsymbol{\mathcal{R}}=\{\mathcal{R}^v\}_{v=1}^N$ are trained with real images to recover the full image from different incomplete views. It is non-trivial to select the appropriate views for completion, which determines what types of frequency-irrelevant information we want to exploit. Since regional consistency, color, and texture have been proven to be distinguishable features for GAN-generated images \cite{liu2020global,hu2021exposing}, we empirically consider three completion tasks: Masked Image Modeling, Gray-to-RGB, and Edge-to-RGB, where the regional, color and textural details are previously missing and restored, respectively. The natural compact distributions of these types of information can be modeled during the completion. The significance of each view is also explored in the experiment.   

\begin{figure}
    \centering
    \includegraphics[width=0.48\textwidth]{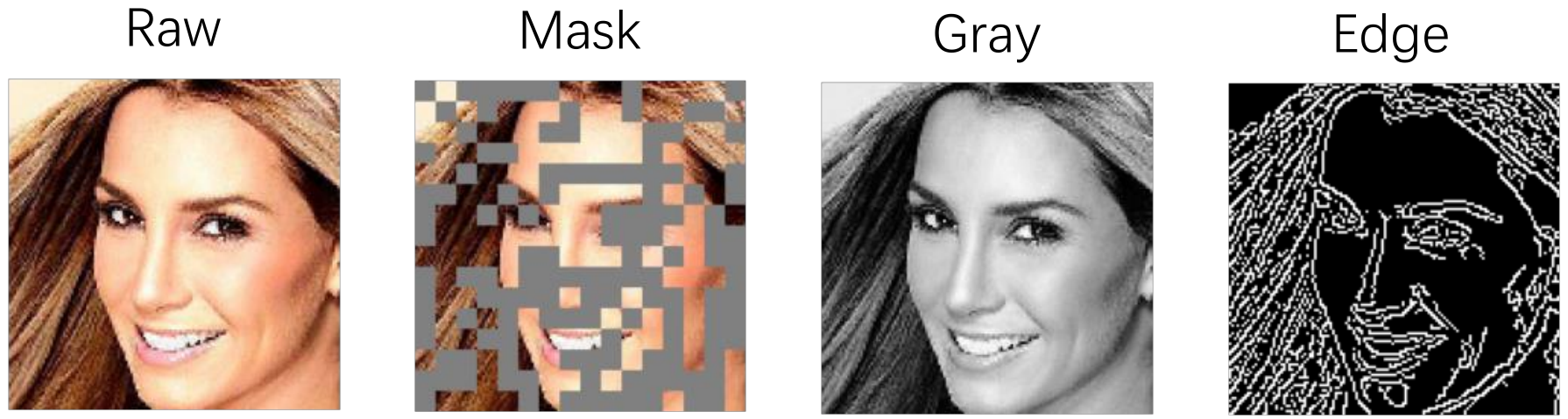}
    \caption{Three incomplete views selected for completion learning.}
    \label{fig:views}
\end{figure}

Masked Image Modeling is an emerging approach for visual representation learning \cite{he2022masked}, which masks a portion of an image and predicts the masked area, and can be leveraged to model the regional consistency of natural images. The masking strategy is that, given an image $X \in \mathbb{R}^{w \times h \times 3}$, we randomly mask $50\%$ non-overlapping patches with a patch size of $(\frac{w}{16}, \frac{h}{16})$. Gray-to-RGB aims to learn color information from real images. We first transform the RGB image into the gray-scale version and then predict the raw RGB pixel values from the gray-scale input. Edge-to-RGB aims to learn textural information from real images. We first extract the binary edge sketch from the RGB image using the Canny edge detector, and then predict the raw RGB pixel values from the edge input. Figure \ref{fig:views} shows an example of different incomplete views. 

\paragraph{Dual-domain constraint.} Given an image $X$ and an incomplete view $X^v$, the completion is formulated as $\Tilde{X}^v=\mathcal{R}^v(X^v)$. The training of $\mathcal{R}^v$ is supervised with a dual-domain constraint, which incorporates a pixel-level regression loss and a frequency loss: 

\begin{equation}
\begin{aligned}
    L_{pix}&=||X - \Tilde{X}^v||_1=||X - \mathcal{R}^v(X^v)||_1, \\
    L_{fre}&=||\mathcal{F}(X) - \mathcal{F}(\Tilde{X}^v)||_2^2.
\end{aligned}
\end{equation}
where $\mathcal{F}(\cdot)$ denotes the 2D FFT. The frequency loss computes the element-wise Euclidean distance between the Fourier spectra of original and restored images, which ensures $\mathcal{R}^v$ to capture the natural, correct frequency property of real images, so as to facilitate the frequency alignment between real and fake images. The optimization function can be therefore denoted as

\begin{equation}
       \min L_{pix}^v + \lambda L_{fre}^v,
        \label{eq-rec-loss}
\end{equation}
where $\lambda$ is the weight to balance different losses.

\subsection{Intra-View Classification Learning}
After training $\mathcal{R}^v$ with real images, both real and fake images are processed by $\mathcal{R}^v$ via the same image completion pipeline to enable the subsequent classification learning. To mine more robust and frequency-irrelevant features from each individual view's pathway, we propose the multi-scale feature concatenation and low-pass residual-guided attention modules, as shown in the orange box in Figure \ref{fig:overview}.

\paragraph{Multi-scale feature concatenation.}
Since $\mathcal{R}^v$ is an encoder-decoder consisting of multiple layers, during the completion, the missing information of the original image is progressively recovered by the stacked decoding layers of $\mathcal{R}^v$. Thus, meaningful features for distinguishing real and fake images are embedded not only in the final output image, but also in the intermediate feature maps of the decoder. To this end, we build a feature pyramid to concatenate the intermediate features at different scales. For a decoder of $\mathcal{R}^v$ with a total of $S$ layers, let $f_s$ be the feature map of the $s$-th layer, the $s$-th feature of the concatenation is computed as: 
\begin{equation}
 \resizebox{.89\linewidth}{!}{$
            \displaystyle
    z_s = \left\{\begin{aligned}
&\operatorname{Conv_3}\left( \operatorname{Concat}\left( \operatorname{Conv_1}(f_s),  \operatorname{Up}(z_{s-1})\right) \right),&s\geq2 \\
&\operatorname{Conv_1}(f_s),&s=1
\end{aligned}
\right.
$}.
\end{equation}
where $\operatorname{Up}(\cdot)$ is an upsampling layer with a scaling factor of $2$ to align the scales between two feature maps; $\operatorname{Conv_1}(\cdot)$ is a $1\times1$ convolutional layer to reduce channel dimensions; $\operatorname{Conv_3}(\cdot)$ is a $3\times3$ convolutional layer to suppress the aliasing effect of upsampling; $\operatorname{Concat}(\cdot)$ indicates the concatenation of two tensors. Finally, the last layer of the feature pyramid $z_S$ is combined with the reconstructed image $\Tilde{X}$ to get the enhanced feature $F$ in the following way:
\begin{equation}
    F= \operatorname{Concat}( \operatorname{Conv_3}(\Tilde{X}),  \operatorname{Conv_3}(z_S))
    \label{eq4}
\end{equation}

\paragraph{Low-pass residual-guided attention.}
The distinguishable features are contained in the restored regional, color, and textural information of the image. Thus, it is possible to leverage the reconstruction residual to provide spatial attention to improve intra-view learning. However, one challenge is that, since the original image $X$ is involved in computing the residual, both stable and unstable features in the original image potentially remain in the residual. As discussed earlier, unstable features that are detrimental to generalization and robustness should be avoided. Prior studies have found that these unstable features are low-level artifacts that mainly cluster in high-frequency components \cite{frank2020leveraging,durall2020watch}. Thus, we propose only using the low-frequency residual to guide the classifier to focus on more stable features. Given an image $X$ and its reconstructed version $\Tilde{X}$, the low-frequency residual is:  
\begin{equation}
    M = |\mathcal{H}(X) - \mathcal{H}(\Tilde{X})|,
\end{equation}
where $\mathcal{H}(\cdot)$ is the first-order low-pass Butterworth filter and $|\cdot|$ is the absolute function. An attention mechanism is then devised to exploit the low-frequency residual. A functional network is used to process $M$ to get the attention map, i.e., $\hat{M} = \mathcal{G}(M)$, where $\mathcal{G(\cdot)}$ consists of a $7\times7$ convolutional layer, an average pooling layer and a sigmoid function. The attention map 
is applied to the enhanced feature $F$ in Eq. \ref{eq4} to obtain the residual-guided feature:
\begin{equation}
    \hat{F} = \hat{M} \otimes \operatorname{Conv_3}(F),
\end{equation}
where $\otimes$ indicates the element-wise multiplication. 

\subsection{Inter-View Classification Learning}
When the intra-view feature enhancement is ready, we can get a set of features $\{\hat{F}^v\}_{v=1}^N$ corresponding to different views. For each view, an independent neural network classifier $\mathcal{C}^v$ is trained on the feature $\hat{F}^v$. Since the features provide view-specific information, the classifiers will learn diverse representations and contribute differently facing the same data instance. To ensure the complementarity and interactivity across different views during training, we propose a self-adaptive cross-view loss fusion strategy. 

\paragraph{Self-adaptive loss fusion.}
The self-adaptive loss fusion strategy aims to combine the losses of different classifiers using adaptive weights, such that the importance of each view-specific representation can be estimated and respected in the final decision. The weights are learned and autonomously adjusted during training. Formally, given a view-specific feature instance $\hat{F}^v$ and the corresponding label $y$ ($y=0$ if the the sample is a real image, otherwise $1$), let $p^v$ be the probability that the sample is fake predicted by $\mathcal{C}^v$. The training of $\mathcal{C}^v$ is supervised by minimizing the cross-entropy loss: 
\begin{equation}
    \min L_{ce}^v := -[y\log(p^v) + (1-y)\log(1-p^v)].
\label{eq-cls-loss}
\end{equation}
The self-adaptive loss fusion strategy can be denoted as a minimization problem with respect to the weights $\boldsymbol{\beta}$:
\begin{equation}
    \min_{\boldsymbol{\beta}}\sum_{v=1}^{N}\beta_{v}^{\tau}L_{ce}^{v} \quad s.t. \quad \boldsymbol{\beta}^\top\boldsymbol{1}=\boldsymbol{1}, \beta_v\geq0,
    \label{eq8}
\end{equation}
where $\tau>1$ is the power exponent parameter to avoid the trivial solution of $\boldsymbol{\beta}$ during the classification. In the inference stage, the decision is made on the average predicted probability of fake ($p^v$) over all classifiers, i.e., $p_{fake}=\frac{1}{3}\sum_{v}p^v$, with a threshold of $0.5$.  

\subsection{Optimization} 
The components of MCCL that require optimization include the parameters of $\{\mathcal{R}^v\}_{v=1}^N$, $\{\mathcal{C}^v\}_{v=1}^N$ and several building blocks for intra-view learning (for simplicity, the latter two are denoted in together as $\{\mathcal{C}^v\}_{v=1}^N$), as well as the self-adaptive loss weights $\boldsymbol{\beta}$. The optimization is performed in the following alternative way:

\paragraph{Update network parameters.} The completion and classification networks with respect to different views are updated independently in parallel. For the view $v$, $\mathcal{R}^v$ and $\mathcal{C}^v$ can be updated sequentially by optimizing the corresponding loss functions Eq. \ref{eq-rec-loss} and Eq. \ref{eq-cls-loss}.
During the optimization, the loss weights $\boldsymbol{\beta}$ are fixed. 

\paragraph{Update loss weights $\boldsymbol{\beta}$.} Next, we fix the parameters of $\{\mathcal{R}^v\}_{v=1}^N$ and $\{\mathcal{C}^v\}_{v=1}^N$, and update $\boldsymbol{\beta}$ by solving Eq. \ref{eq8}. To satisfy the constraints, the Lagrangian function of Eq. \ref{eq8} is
    \begin{equation}
        \mathcal{L(\boldsymbol{\beta}, \zeta)} = \sum_{v=1}^{N}\beta_{v}^{\tau}L_{ce}^{v}-\zeta(\sum_{v=1}^N\beta_v-1)
        \label{eq10}
    \end{equation}
    where $\zeta$ is the Lagrange multiplier. By derivation of Eq. \ref{eq10} with respect to $\beta_{v}$ and $\zeta$, the optimal solution of Eq. \ref{eq8} is:
    \begin{equation}
        \beta_{v} = (L_{ce}^v)^\frac{1}{1-\tau}/{\sum_{n=1}^{N}(L_{ce}^n)^\frac{1}{1-\tau}}
    \end{equation}

\section{Experiments} 

\subsection{Datasets}
\paragraph{Real images.} Our experiments are conducted on facial images, given that human faces are the main target of deepfakes. We choose the large-scale facial image dataset CelebA \cite{liu2015faceattributes} and its high-quality version CelebA-HQ \cite{karras2018progressive} to perform evaluations at different resolutions. The CelebA images have a resolution of $128\times128$ and the CelebA-HQ images have a resolution of $1024 \times 1024$. 

\paragraph{GAN-generated images.} A total of six popular GAN types are considered: ProGAN \cite{karras2018progressive}, CramerGAN \cite{bellemare2017cramer}, SNGAN \cite{miyato2018spectral}, MMDGAN \cite{li2017mmd}, StyleGAN \cite{karras2019style} and StyleGAN2 \cite{karras2020analyzing}. In the low-resolution setting, we follow the setting in \cite{yu2019attributing}, using the pre-trained ProGAN, CramerGAN, SNGAN, and MMDGAN models \footnote{\url{https://github.com/ningyu1991/GANFingerprints}} to generate fake faces. All the four GANs are pre-trained with CelebA. In the high-resolution setting, we adopt the dataset released by \cite{he2021beyond} \footnote{\url{https://github.com/SSAW14/BeyondtheSpectrum}}, which includes images generated by ProGAN, StyleGAN, and StyleGAN2. Note that the ProGAN and StyleGAN are pre-trained with CelebA-HQ, while the StyleGAN2 with another facial image dataset FFHQ \cite{karras2019style}. Since FFHQ has a larger diversity in terms of facial attributes compared with CelebA-HQ, StyleGAN2 is included for cross-domain evaluation. Table \ref{tab:dataset} shows the details of the dataset setting. 

\begin{table}[htbp]
\renewcommand\arraystretch{1}
  \centering
    \resizebox{0.48\textwidth}{!}{
    \setlength\tabcolsep{1mm}{
    \begin{tabular}{lccccc}
    \toprule
    \multicolumn{1}{l|}{128x128} & CelebA & ProGAN & CramerGAN & SNGAN & MMDGAN \\
    \midrule
    \multicolumn{1}{l|}{Training } & 60k & 60k & $-$   & $-$   & $-$ \\
    \multicolumn{1}{l|}{Test} & 6k & 6k & 6k & 6k & 6k \\
    \midrule
    \multicolumn{1}{l|}{1024x1024} & CelebA-HQ & ProGAN & StyleGAN & StyleGAN2$^*$  \\
    \midrule
    \multicolumn{1}{l|}{Training } & 25k & 25k & 25k & $-$   &  \\
    \multicolumn{1}{l|}{Test} & 2.5k & 2.5k & 2.5k & 2.5k &  \\
    \midrule
    \multicolumn{6}{l}{* Pre-trained in a different real image dataset FFHQ} \\
    \end{tabular}}}%
\caption{The details of the experimental dataset setting.}
    \label{tab:dataset}%
\end{table}%

\subsection{Implementation Details}
The restorers and classifiers are implemented based on U-Net \cite{ronneberger2015u} and Xception \cite{chollet2017xception}, respectively. The U-Net we use has five skip connection blocks (i.e., $S=5$), and their output feature maps are employed to build the feature concatenation. We train the whole framework with a batch size of 80 using the Adam optimizer \cite{kingma2015adam}. The initial learning rate is 1e-3, and we reduce it to half after every ten epochs. $\tau$ in Eq. \ref{eq8}, and $\lambda$ in Eq. \ref{eq-rec-loss} are empirically set to 4 and 10, respectively. We also use random Gaussian noise, color jitter, and blurring for data augmentation on the restorer side. 

\begin{table}[htbp]
\renewcommand\arraystretch{1}
  \centering

  \resizebox{0.48\textwidth}{!}{
        \begin{tabular}{l|cccccccc}
    \toprule
          & \multicolumn{2}{c}{ProGAN} & \multicolumn{2}{c}{CramerGAN} & \multicolumn{2}{c}{SNGAN} & \multicolumn{2}{c}{MMDGAN} \\
\cmidrule{2-9}          & Acc.   & A.P.  & Acc.   & A.P.  & Acc.   & A.P.  & Acc.   & A.P. \\
    \midrule
    GAN-FP & 99.5  & 99.8  & 52.1  & 55.4  & 53.6  & 70.4  & 48.2  & 53.2 \\
    2d-DCT & 98.9  & 99.1  & 70.2  & 67.1  & 61.9  & 73.5  & 56.0  & 73.1 \\
    DA    & 99.5  & 99.9  & 72.1  & 78.3  & 63.1  & 70.0  & 54.7  & 71.7 \\
    FLP   & 95.1  & 98.3  & 81.3  & 81.7  & \textbf{83.6}  & 80.1  & 70.2 & 82.0 \\
    SRR   & \textbf{100.}  & \textbf{100.}  & 88.2  & \textbf{95.1 } & 70.3  & 81.5  & 77.7  & 84.5 \\
    MCCL (Ours)  & \textbf{100.}  & \textbf{100.}  & \textbf{91.1}  & 89.2  & 80.2  & \textbf{83.3}  & \textbf{85.4}  & \textbf{86.1} \\
    \bottomrule
    \end{tabular}
    }
  \caption{The results of cross-GAN detection in the $128\times128$ setting. \textbf{Bold} indicates the best score in each column.}
    \label{tab:lr_crossgan}%
\end{table}%

\begin{table*}[htbp]
\renewcommand\arraystretch{1}
  \centering

  \resizebox{0.7\textwidth}{!}{

    \begin{tabular}{l|cccc|cccc|cccc}
    \toprule
    \multicolumn{1}{r}{} & \multicolumn{4}{c|}{Within-distribution} & \multicolumn{4}{c|}{Cross-GAN} & \multicolumn{4}{c}{Cross-GAN \& Cross-domain} \\
\cmidrule{2-13}    \multicolumn{1}{c}{} & \multicolumn{2}{c}{P$\rightarrow$P} & \multicolumn{2}{c|}{S$\rightarrow$S} & \multicolumn{2}{c}{P$\rightarrow$S} & \multicolumn{2}{c|}{S$\rightarrow$P} & \multicolumn{2}{c}{P$\rightarrow$S2} & \multicolumn{2}{c}{S$\rightarrow$S2} \\
    \multicolumn{1}{c}{} & Acc.   & A.P.  & Acc.   & A.P.  & Acc.   & A.P.  & Acc.   & A.P.  & Acc.   & A.P.  & Acc.   & A.P. \\
    \midrule
    GAN-FP & 99.9  & 99.9  & 99.4  & 99.6  & 51.1  & 71.0  & 49.3  & 68.8  & 44.3  & 47.6  & 48.0  & 46.9 \\
    2d-DCT & 99.9  & 99.9  & 99.8  & 99.9  & 90.1  & 91.5  & 93.0  & 92.1  & 62.1  & 60.0  & 93.8  & 90.2 \\
    DA    & 97.6  & 96.6  & 98.3  & 97.8  & 73.2  & 87.7  & 78.1  & 73.1  & 66.1  & 79.1  & 80.7  & 84.4 \\
    FLP   & 98.9  & 99.0  & 99.1  & 98.9  & 95.0  & 97.1  & 94.3  & 86.3  & 80.8  & 88.0  & 92.4  & 93.1 \\
    SRR   & \textbf{100.}  & \textbf{100.}  & 99.9  & 99.9  & \textbf{99.1}  & \textbf{99.4}  & 98.2  & 98.1  & 88.2  & 80.3  & 91.5  & 91.1 \\
    MCCL (Ours)  & \textbf{100.}  & \textbf{100.}  & \textbf{100.}  & \textbf{100.}  & 98.1  & 95.5  & \textbf{99.2}  & \textbf{98.9}  & \textbf{95.3}  & \textbf{90.0}  & \textbf{97.7}  & \textbf{96.0} \\
    \bottomrule
    \end{tabular}
    }%

  \caption{The results of cross-GAN detection in the $1024\times1024$ setting. \textbf{Bold} indicates the best-in-column. P, S, S2 are short for ProGAN, StyleGAN and StyleGAN2, respectively. The right and left sides of $\rightarrow$ indicate the training and test sets, respectively.}
  \label{tab:hr_crossgan}%
\end{table*}%
\begin{table*}[htbp]
 \renewcommand\arraystretch{1}
  \centering

  \resizebox{1\textwidth}{!}{

    \begin{tabular}{lcc|cc|cc|cc|cc|cc|cc|cc|cc}
    \toprule
          & \multicolumn{2}{c|}{Clean} & \multicolumn{2}{c|}{Blurring} & \multicolumn{2}{c|}{Cropping} & \multicolumn{2}{c|}{Compression} & \multicolumn{2}{c|}{Noise} & \multicolumn{2}{c|}{Mix} & \multicolumn{2}{c|}{FGSM} & \multicolumn{2}{c|}{PGD} & \multicolumn{2}{c}{SDN} \\
\cmidrule{2-19}          & Acc.   & A.P.  & Acc.   & A.P.  & Acc.   & A.P.  & Acc.   & A.P.  & Acc.   & A.P.  & Acc.   & A.P.  & Acc.   & A.P.  & Acc.   & A.P.  & Acc.   & A.P. \\
    \midrule
    GAN-FP & 99.5  & 99.8  & 49.6  & 67.4  & 44.9  & 77.5  & 8.7   & 45.8  & 9.0   & 49.1  & 19.3  & 66.6  & 11.1  & 15.5  & 8.1   & 22.1  & 13.4  & 45.0  \\
    2d-DCT & 98.9  & 99.1  & 60.4  & 77.7  & 80.5  & 76.1  & 67.4  & 80.2  & 46.7  & 74.3  & 61.3  & 61.8  & 34.0  & 45.3  & 23.1  & 41.3  & 21.8  & 56.1  \\
    DA    & 99.5  & 99.9  & 83.2  & \textbf{98.9 } & 51.8  & 64.1  & 84.0  & \textbf{97.3 } & 74.3  & 80.2  & 85.5  & 91.0  & 43.4  & 66.7  & 40.1  & 54.4  & 56.7  & 67.0  \\
    FLP   & 95.1  & 98.3  & 96.1  & 90.2  & 71.6  & 77.0  & 80.3  & 74.3  & 90.9  & 91.1  & 84.7  & 89.9  & 56.1  & 60.7  & 49.4  & 67.0  & 43.2  & 60.1  \\
    SRR   & \textbf{100.} & \textbf{100.} & 92.1  & 93.0  & 97.9  & 96.1  & 90.7  & 93.3  & 92.0  & 88.8  & 89.6  & 90.6  & 67.1  & 75.2  & 64.8  & 77.1  & 87.2  & 91.1  \\
    MCCL (Ours)  & \textbf{100.} & \textbf{100.} & \textbf{96.4 } & 98.5  & \textbf{98.2 } & \textbf{99.1 } & \textbf{93.8 } & 96.9  & \textbf{94.7 } & \textbf{94.4 } & \textbf{91.3 } & \textbf{94.4 } & \textbf{81.6 } & \textbf{80.3 } & \textbf{81.3 } & \textbf{81.9 } & \textbf{93.2 } & \textbf{95.6 } \\
    \bottomrule
    \end{tabular}%
    }%
  \caption{The results of robustness against 8 perturbation methods. \textbf{Bold} indicates the best score in each column. }
  \label{tab:robustness}%
\end{table*}%

\subsection{Baseline Detection Models}
We compare MCCL with two normal detectors that exploit a CNN to extract and learn features directly: the image-domain detector using GAN fingerprints (GAN-FP) \cite{yu2019attributing} and the frequency-domain detector based on 2D DCT coefficients (2d-DCT) \cite{frank2020leveraging}. We also compare three state-of-the-art generalized and robust detection methods: the data augmentation-based method (DA) \cite{wang2020cnn}, the frequency-level perturbation (FLP) \cite{jeong2022frepgan}, and the super-resolution re-synthesis (SRR) \cite{he2021beyond}, each of which has specific strategies to improve robustness. The detection performance is evaluated by the classification accuracy (Acc.) and the average precision score (A.P.) commonly used in related studies \cite{wang2020cnn,jeong2022frepgan}.

\subsection{Results of Generalization}
\paragraph{Low-resolution setting.} We train the detection model with the CelebA images and the corresponding $128\times128$ ProGAN images and test it with the ProGAN, CramerGAN, SNGAN, and MMDGAN images to evaluate the cross-GAN generalization ability. The results compared to five baselines are listed in Table \ref{tab:lr_crossgan}. We conclude that: 1) The normal detectors, GAN-FP and 2d-DCT, are highly accurate for within-distribution detection but generalize poorly for cross-GAN detection, implying the risk of overfitting unstable features; 2) The other four detectors with specific strategies all get the cross-GAN detection performance improved. Our method MCCL achieves the best or second-best results spanning over all GANs, thanks to the novel multi-view representation enriching more robust, frequency-irrelevant features. 

\paragraph{High-resolution setting.} We follow the setting in \cite{he2021beyond}: we train a ProGAN detector and a StyleGAN detector independently using CelebA-HQ and the corresponding GAN images, and test both on ProGAN, StyleGAN, and StyleGAN2 test samples to evaluate the within-distribution, cross-GAN and cross-domain performances, respectively. The results are summarized in Table \ref{tab:hr_crossgan}. All detectors perform well in within-distribution detection. In the cross-GAN group, 2d-DCT becomes more robust compared with its low-resolution performance. The reason may be that with the resolution increasing, stable low-frequency features are naturally enriched, which can be easier captured by the classifier trained directly on spectra. We can also see that FLP, SRR, and MCCL improve more significantly than DA in this group because they reduce classifiers' dependency on unstable frequency features in a learnable way. Regarding the cross-GAN \& cross-domain group, which is the most challenging, our method remarkably outperforms all baseline methods in both sub-groups, indicating great applicability to difficult detection scenarios. 

\begin{figure*}[htbp]
    \centering
    \includegraphics[width=1\textwidth]{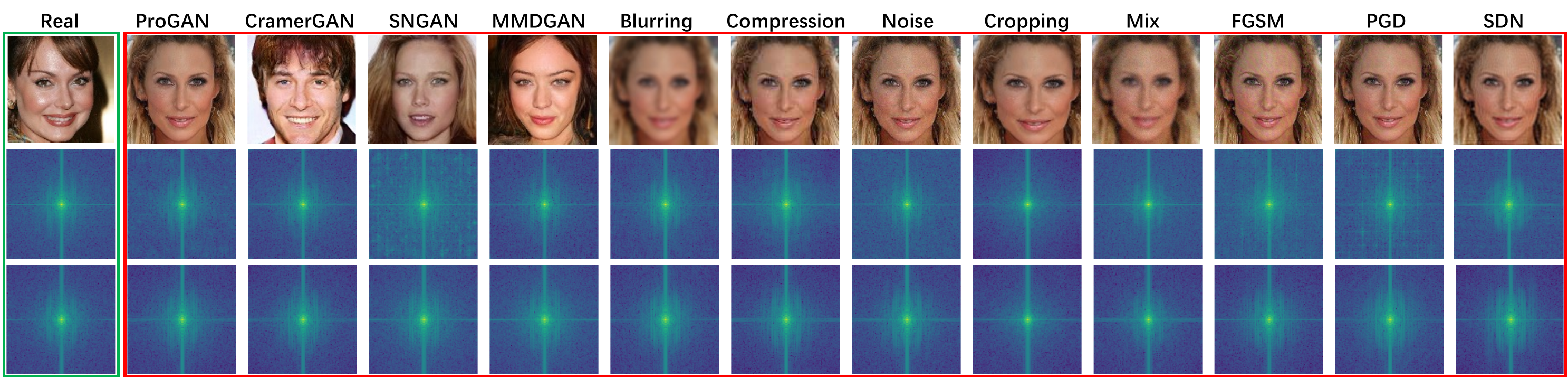}
    \caption{The visualization of real and different GAN-generated and perturbed fake samples (the 1st row) and the average FFT spectra before (the 2nd row) and after (the 3rd row) the Edge-to-RGB completion.}
    \label{fig:visual}
\end{figure*}

\subsection{Results of Robustness}
We evaluate the robustness against perturbations using the $128\times128$ ProGAN detectors. We train detectors with the CelebA and ProGAN images, and test them with perturbed ProGAN samples. Unlike prior work mainly concerning common image manipulations \cite{frank2020leveraging,yu2019attributing,wang2020cnn}, we investigate a broader range of perturbations: (1) Common manipulations, including Blurring, Cropping, Compression, Noising and a mix of all. We follow the setting in \cite{frank2020leveraging} to created the perturbations; (2) Adversarial perturbations including FGSM \cite{goodfellow2014explaining} and PGD \cite{madry2018towards}. The adversarial examples are crafted based on a vanilla Xception detector with the noise amount $\epsilon = 8/255$; and (3) Spectrum Difference Normalization (SDN) \cite{dong2022think}, an attack specific to GAN-generated images that calibrates the spectra of fake images according to real images.

An example of samples modified by different perturbations is shown in Figure \ref{fig:visual}. Table \ref{tab:robustness} shows the results. Since most perturbations significantly modify the original frequency distribution of fake samples, the performance of normal detectors degrades rapidly, while the other four are relatively more resistant given the suppression of frequency overfitting. Among the four robust methods, our method achieves the best results regarding all perturbations except for the A.P. scores for blurring and compression. It is worth noting that, when confronted with much more challenging perturbations such as the adversarial attacks FGSM and PGD and the specific attack SDN, our method obtains Acc. and A.P. scores that are notably higher than other baselines.

\paragraph{Summary.} We here discuss the superiority of MCCL over the other three robust detection methods DA, FLP and SRR. DA and FLP augment the frequency distributions in the training set by image processing or adversarial perturbation to improve the generalization to unknown frequency patterns. However, only augmentation is insufficient as it cannot cover all GAN-specific frequency patterns. Unlike the frequency-level augmentation, MCCL seeks to eliminate the frequency distributional gaps between real and fake images via image completion, which can enforce the subsequent classifiers to learn more stable, frequency-irrelevant features. SRR employs a super-resolution process to model real image distribution, which can be regarded as a single-view image completion task. MCCL outperforms SRR because of the multi-view representation and several novel learning strategies, which allow stable and robust feature representation from diverse types of information.

\subsection{Analysis and Discussion}

\begin{table}[htb]
\renewcommand\arraystretch{1}
  \centering

  \resizebox{0.48\textwidth}{!}{

  \setlength{\tabcolsep}{1mm}{
    \begin{tabular}{ccc|cccccc}
    \toprule
          &       & \multicolumn{1}{r}{} & \multicolumn{2}{c}{Within-distribution} & \multicolumn{2}{c}{Cross-GAN} & \multicolumn{2}{c}{Cross-perturbation} \\
\cmidrule{4-9}          &       & \multicolumn{1}{r}{} & Acc.  & A.P.  & mAcc. & mA.P. & mAcc. & mA.P. \\
    \midrule
    Masked   & Gray  & Edge  &       &       &       &       &       &  \\
     $\checkmark$ &  &       & 99.6  & 99.0  & 82.1  & 81.3  & 83.3  & 87.5  \\
      &   $\checkmark$    &       & 99.1  & 99.3  & 67.2  & 72.1  & 71.1  & 77.0  \\
     &  &  $\checkmark$     & 100.0  & 100.0  & 78.9  & 83.4  & 88.8  & 90.1  \\
    \midrule
    MFC   & LRA   & ALF   &       &       &       &       &       &  \\
     &       &       & 93.7  & 97.0  & 75.2  & 73.1  & 81.4  & 74.9  \\
    $\checkmark$ &       &       & 95.1  & 97.3  & 79.1  & 73.0  & 86.8  & 79.9  \\
    $\checkmark$ & $\checkmark$ &       & 97.5  & 98.9  & 81.2  & 76.1  & 90.5  & 82.3  \\
    \midrule
    \multicolumn{3}{c|}{Final$\quad$version} & 100.0  & 100.0  & 85.6  & 86.2  & 91.3  & 92.6  \\
    \bottomrule
    \end{tabular}}}
      \caption{The results of ablation studies with different views or modules. MFC: Multi-scale Feature Concatenation. LRA: Low-pass Residual-guided Attention. ALF: Adaptive Loss Fusion.}
        \label{tab:ablation}
\end{table}%
\begin{figure}[htbp]
    \centering
    \includegraphics[width=0.48\textwidth]{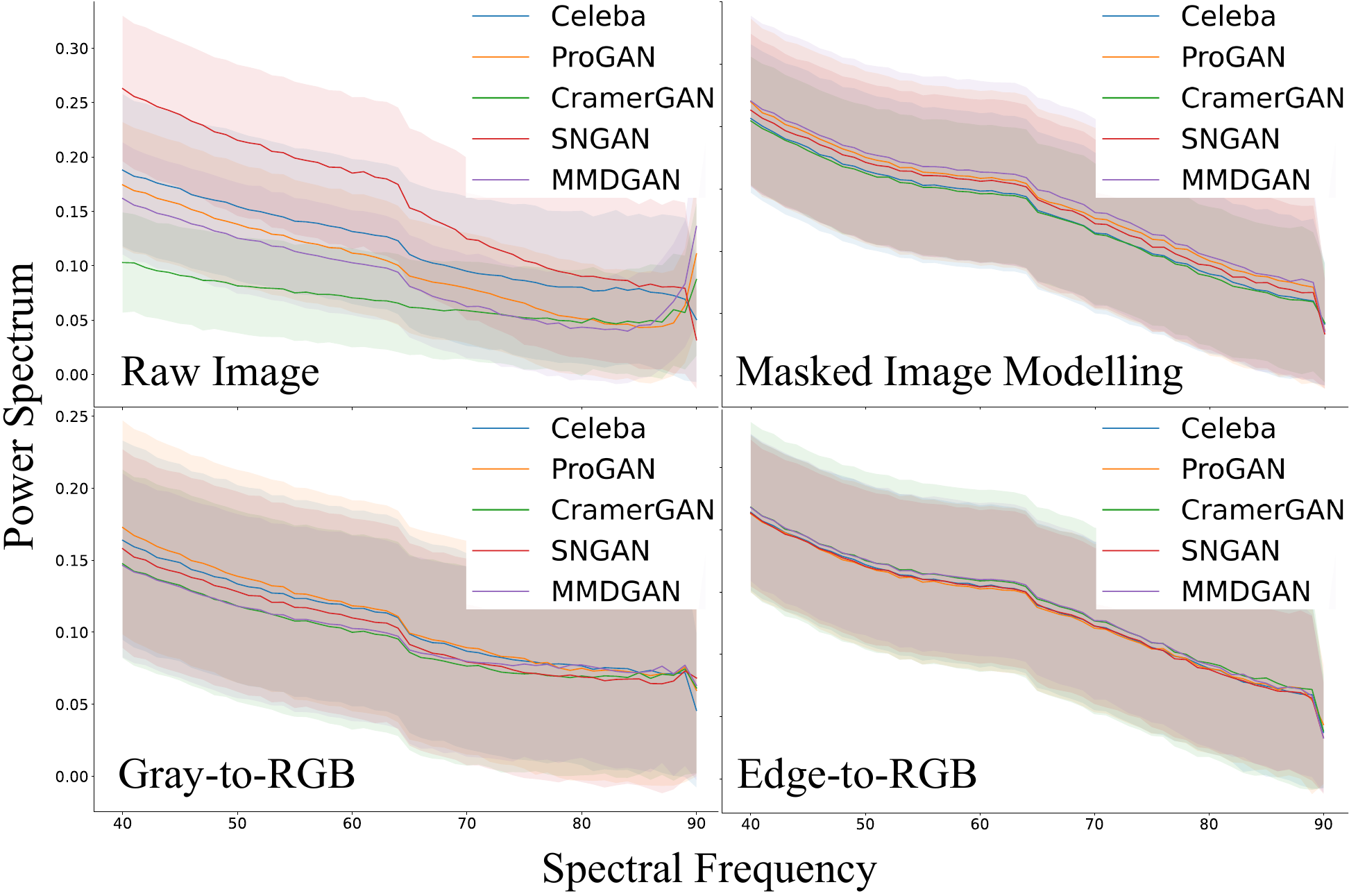}
    \caption{The spectral distributions of real images and fake images generated by different GANs before and after completion. }
    \label{fig:frequency} 
\end{figure}
\paragraph{Ablation study.} Two ablation studies are performed based on the $128\times128$ ProGAN detectors to show the significance of individual views and the effects of different modules . We report the average Acc. (mAcc.) and A.P. (mA.P.) scores for cross-model and cross-perturbation performance, as shown in Table \ref{tab:ablation}. Regarding different view settings, the gray view leads to a relatively weaker robustness than the other two views, indicating that color information is less distinguishable than regional consistency and texture. The combination of all views (i.e., the final version in Table \ref{tab:ablation}) outperforms all individual views, confirming the effectiveness of multi-view representation and the capability of MCCL to capture and fuse different types of view-specific features for robust detection. Additionally, we emphasize that MCCL is fully flexible and extensible in view configuration, which can incorporate more quantities and types of views to enable stronger feature representations. Regarding different module settings, with more modules activated, the learning capacity of the model improves, enabling more effective feature representation. 

\paragraph{Frequency analysis.} One advantage of multi-view completion representation is that it helps reduce the classifier's reliance on unstable frequency patterns by aligning the frequency distributions between real and fake images. It works because the unstable frequency artifacts of fake samples are prior removed in the incomplete views, and then the missing frequency pattern is reconstructed and calibrated according to real images by the restorer pre-trained with real images. We provide a spectral analysis to confirm the frequency alignment by estimating the spectral distributions using the azimuthal integration over each radial frequency of the center-shifted FFT spectrum \cite{durall2020watch}. Figure \ref{fig:frequency} shows the distribution curves of real images and images generated by different GANs before and after completion. The original distributions differ significantly between real and fake images and between different GANs. As a result, a CNN detector may easily overfit one specific frequency pattern that may not generalize to another. After the view-to-image completions, the spectral distributional gaps have become much closer. The alignment is more thorough in the edge-to-RGB completion than in the other two because edge sketches remove far more information from the original image than masked and gray views. These frequency-aligned training samples will force the classifier to focus on more stable, general, and frequency-insensitive features. Figure \ref{fig:visual} provides a visualization of the averaged FFT spectra of different fake samples before and after the edge-to-RGB completion. The differences between real images and all types of fake images become visibly smaller after completion. 

\section{Conclusion}
To overcome the generalization and robustness issues in GAN-generated image detection, we propose a novel framework incorporating multi-view completion learning and cross-view classification learning for a robust feature representation. Numerous experiments with varying cross-resolution, cross-GAN, and cross-perturbation settings validate the outperforming generalization and robustness of our proposed framework compared with the current state-of-the-art detectors. We also confirm the effect of reducing frequency reliance in deepfake detection, offering a potential route for future designs of robust deepfake detection.  

\section*{Acknowledgements}
This work is supported by an ARC Discovery Project DP230100246 from the Australian Research Council Australia.
\bibliographystyle{named}
\bibliography{ijcai23}

\end{document}